\documentclass[prb,twocolumn,amsmath]{revtex4}
\usepackage{graphicx}
\newcommand{\be}{\begin{equation}}
\newcommand{\ee}{\end{equation}}
\newcommand{\<}{\langle}
\renewcommand{\>}{\rangle}

\begin{document}

\title{ Scissors Mode and dichroism in an anisotropic crystal }



\author{Keisuke Hatada$^{1}$, Kuniko Hayakawa$^{1, 2}$, Fabrizio Palumbo$^{1}$}

\address{$^{1}$INFN Laboratori Nazionali di Frascati, c.p. 13, I-00044 Frascati, Italy \\
$^{2}$Dipartimento di Biologia, Universita ``Padova'', Via Ugo Bassi 58b, 35121 Padova, Italy \\
{\small e-mail: {\tt palumbof@lnf.infn.it}}}

\begin{abstract}

We suggest that in an anisotropic crystal there should be a new mechanism of dichroism 
related to a scissors mode, a kind of
excitation observed in several other many-body systems. Such an effect should
be found in crystals, amorphous systems and also metallo-proteins.
Its signature is a strong magnetic dipole transition amplitude,
which is a function of the angle between the momentum of the photon
and the anisotropy axis of the cell. 

\end{abstract}
\maketitle

The so-called scissors mode is a collective excitation,
where two systems can rotate with respect to each other conserving their shape.
It was predicted to occur in deformed atomic nuclei~\cite{loiudice},
where protons and neutrons are assumed to form two distinct bodies
to be identified with the blades of the scissors. Their relative rotation
generates a magnetic dipole moment which
provides the signature of the mode.

In this paper we suggest that a similar excitation should exist as well
in all systems where there is a deformed atom in a distorted cell,
like crystals, for instance perovskites,
and with the qualifications discussed at the end, also amorphous systems
and metallo-proteins. In all these systems the scissors mode should give
rise to a new mechanism of dichroism,
since the atom can rotate only around the anisotropy axis, and therefore
 cannot absorb a photon with momentum perpendicular to it, as shown in the Figure.

The investigation of this mechanism of dichroism has several motivations. The most obvious is a direct
determination of the distortion parameters. In particular for metallo-proteins this could
provide a new approach to the study of their structure.

A second motivation concerns theoretical calculations of dispersive effects
because these will be dominated
by the scissors mode in a channel with its quantum numbers.

A last reason of general interest is in the comparison with similar excitations
predicted or observed in other systems.

Indeed, after Richter and coworkers~\cite{bohle} discovered the scissors mode
in $^{156}Gd $, it was systematically investigated both
experimentally~\cite{rich} and theoretically~\cite{loiudice1} in atomic nuclei.
Later on it was predicted to occur in metal clusters~\cite{lipparini}, quantum dots~\cite{serra},
Bose-Einstein~\cite{gueri} and Fermi~\cite{minguzzi} condensates.
In all these systems, like in the present case, one of the blades of the scissors must be identified
with a moving part (i.e. the valence electrons in metal clusters and
quantum dots or the atom cloud in Bose condensates) and the other one
with a structure at rest (the ions in metal clusters or the trap
in quantum dots and Bose-Einstein condensates). There is one exception
in Bose-Einstein condensates, where the simultaneous condensation of
two different species of atoms has been obtained~\cite{ingu}, thus realizing
a situation closer to the nuclear case. Until now, as far as we know,
among all these systems the scissors excitation has been discovered
only in Bose-Einstein condensates~\cite{marago}, and are under experimental
investigation in metal clusters~\cite{portales}.

Also in these cases the study of the scissors mode provides several pieces of distinctive
information. In Nuclear Physics it is related to the superconductivity
of deformed nuclei, in Bose-Einstein condensates it provides a signature
of superfluidity, in metal clusters it is predicted to be responsible
for paramagnetism.

In the present case numerical estimates are easy for ionic bindings, and as an example we consider
the distorted crystal LaMnO$_3$, which has a perovskite
structure and has recently attracted attention due to colossal magnetoresistance
effects. We find that the lowest excited state has an energy
of about 4 eV with a $M1$ transition amplitude of 1.5 a.u. We think that
these values can be taken as an order of magnitude also for amorphous systems
and metallo-proteins which have covalent binding.

We study the motion of the atom in the cell frame of reference.
We assume the $z$-axis parallel to the anisotropy axis.
We do not need to specify the directions of the $x$-, $y$-axes
because we assume axial symmetry. We will also use the intrinsic frame of
reference of the atom, with the $\zeta$-axis parallel to its symmetry axis.
We call $\theta$ the angle between the $\zeta$- and the $z$-axis,
and we denote the components of ${\hat \zeta}$ in the cell frame according to
$ \zeta_{x} = \sin\theta\,\cos\varphi$, $\zeta_{y}
= \sin\theta\,\sin\varphi$ and $ \zeta_{z} = \cos\theta $.

The potential energy of the atom depends only on $\theta$, and we take it to be
\begin{equation}
  V = \frac{1}{2}\,C\,\sin^2\theta.
\end{equation}
We identify two of the Euler angles of the intrinsic frame
with $\varphi$ and $\theta$, the third angle $\gamma$ defining
the orientation of the atom around the $\zeta$-axis. We assume
the kinetic energy of the atom to be that of a rigid rotator.

In quantum mechanics rotations around the symmetry axis are forbidden,
and therefore the wave functions must satisfy the constraint
$ {\hat {\boldsymbol {\zeta}}}\,\cdot\,{\bf J}
= -i \, {\partial \over \partial \gamma} = \,0 $,
where ${\bf J}$ is the angular momentum the atom.
Its components in the cell frame become then those
of a point particle so that
\begin{equation}
  T = \,\frac{\hbar^2}{2I_{1}}\,\biggl(\, - \frac{\partial^2}{\partial\,\theta^2}
  - \cot\,\theta\,\frac{\partial}{\partial\,\theta} +\frac{1}{\sin^2\,\theta}
  \,J_{z}^2 \,\biggr),
\end{equation}
where $I_1$ is the moment of inertia along the one intrinsic axis.
There remains another constraint. Indeed the two orientations
of the atomic symmetry axis cannot be distinguished from one another,
so that the admissible wave functions must be symmetric under the operation
\begin{eqnarray}
  r\,:\,
      \theta \, \rightarrow \,\pi - \theta, \,
      \varphi \, \rightarrow \,\pi + \varphi.
\end{eqnarray}
Needless to say, the Hamiltonian is symmetric under this operation.
The scalar product for such a system remains defined by the measure
of integration over the Euler angles, which for functions independent
of $\gamma$ is 
\begin{equation}
  \<\,\psi_i\,|\,\psi_f\,\> = \int_{0}^{2\pi}d\varphi\int_{0}^{\pi}d\theta
  \sin\theta\,\psi_i^{*}(\theta,\varphi)\,\psi_f(\theta,\varphi).
\end{equation}
Exploiting the $r$-symmetry we will write the matrix elements in the form
\begin{eqnarray}
  \<\,\psi_i\,|\,{\it O}\,|\,\psi_f\,\> &=& \int_{0}^{2\pi}d\varphi\int_{0}^{\pi/2}
  d\theta\,\sin\theta \nonumber \\
  &&\hspace{-30mm}\times\,\psi_i^{*}(\theta,\varphi)\,
  [\,{\it O}(\varphi,\theta)+\,{\it O}\,(\,\pi+\varphi,\,\pi-\theta\,)\,]
  \,\psi_f(\theta,\varphi).
\end{eqnarray} 
Notice that accordingly the wave functions are normalized to $ { 1\over 2} $.

The exact eigenfunctions and eigenvalues are known \cite{abramowitz},
but in the cases of physical interest the quadratic approximation is sufficient
\begin{eqnarray}
  T &=& \,\frac{\hbar^2}{2I_{1}}\,\biggl(\,- \frac{\partial^2}{\partial\,\theta^2}
  - \frac{1}{\theta}\,\frac{\partial}{\partial\,\theta}
  + \frac{1}{\theta^2}\,J_{z}^2 \,\biggr), \quad 0\le \theta \le \pi \nonumber \\
  V &=& 
  \left\{ 
    \begin{array}{rl}
      &1/2\,C\,\theta^2 \hspace{12mm},\quad\,0\,\le\,\theta\,\le\,\pi/2,\\
      &1/2\,C\,(\,\pi - \theta\,)^2\,\,\,,\quad\,\pi/2\,<\,\theta\,\le\,\pi.
    \end{array}
  \right.
  \label{eqn:kin}
\end{eqnarray}
We recognize that this Hamiltonian, in each of the $\theta$-regions above,
is exactly that of a two-dimensional harmonic oscillator,
provided we identify $\theta$ with the polar radius. As shown by the
following estimates, the fall off of the wave function is so fast
that we can extend without any appreciable error the integral over $\theta$
up to infinity.

In the quadratic approximation, in the region $0\,\le\,\theta\,\le\,\pi/2$
the eigenfunctions are
\begin{eqnarray}
  \psi_{n,m}(\theta,\varphi) &=& \frac{1}{\sqrt{2\pi}}\,\,
  {e^{im\varphi}}\chi_{ n, |m| }(\theta)
\end{eqnarray}
where
\begin{eqnarray}
  \chi_{ n, |m| }(\theta) = \nu_{ n, |m| }\biggl(\frac{\theta}{\theta_{0}}\biggr)^{|m|}
  {\rm exp}\biggl(-\frac{\theta^2}{2\theta_{0}^2}\biggr)L^{(|m|)}_{n}
  \biggl(\frac{\theta^2}{\theta_{0}^2}\biggr).
\end{eqnarray}
$L^{(|m|)}_{n}$ are Laguerre polynomials \cite{abramowitz} and
\begin{eqnarray}
  \nu_{ n, |m| }=\sqrt{\frac{n!}{\theta_{0}^2\,\Gamma\,(\,n+|m|+1\,)}},
  \,\,\,\theta_0^2=\frac{\hbar}{\sqrt{I_1\,C}}.
\end{eqnarray}
Because of the $r$-symmetry in the region $\pi/2\,<\,\theta\,\le\,\pi$, we have
$ \chi_{ n, |m| }(\pi- \theta)=(-1)^m \, \chi_{ n, |m| }( \theta) $.
The eigenvalues are
\begin{equation}
  E_{n,m} = \hbar\omega\,(\,2n + |m| +1\,)\,,\hspace{5mm}(\,n = 0,\,1,\,2\,...)
\end{equation}
where 
\begin{eqnarray}
  \omega = \sqrt{\frac{C}{I_1}}.
\end{eqnarray}
The hamiltonian~(\ref{eqn:kin}) is exactly the intrinsic
hamiltonian which appeared in the nuclear model.

The first excited states have $n=0$, $m=\pm 1$, and as we will see, are the only states
strongly coupled to the ground state by electromagnetic radiation.
As shown in the Figure, they describe
the precession of the atom at an angle $\overline{\theta} \sim \theta_0$
around the axis of the cell. Due to the small value of $\theta_0$
the atom is essentially polarized. This, as explained below,
has an important consequence on the experimental measurements.

Now we evaluate the $M1$ and $E2$ electromagnetic transition amplitudes.
We will find that the $M1$-amplitude is large, being proportional to $1/\theta_0$,
while the $E2$ one is proportional to $\theta_0$ and therefore small.

The intrinsic magnetic dipole moment of the atom is,
\begin{equation}
  {\cal M}\,(\,M1\,,\,\mu\,)\, = -\frac{1}{2\,c}\,\int\,d{\bf r}\,
  {\bf j}\,(\,{\bf r}\,)\,\cdot\,{\bf r}\wedge\nabla\,
  \biggl(r\,Y_{1\mu}\,(\,{\hat {\bf r}}\,)\biggr).
\end{equation}
The atomic electric current
\begin{equation}
  {\bf j}\,(\,{\bf r}\,)\, = \frac{e}{m_e}\,\rho\,(\,{\bf r}\,)\,
  {\bf p}\,(\,{\bf r}\,)
\end{equation}
depends on the electric charge $e$, the mass $m_e$,
the momentum ${\bf p}\,(\,{\bf r}\,)$ of the electrons and the atomic density $\rho\,(\,{\bf r}\,)$.
The intrinsic atomic magnetic moment ${\cal M} $ can be rewritten
\begin{eqnarray}
  {\cal M}\,(\,M1\,,\,\mu\,)\, = \frac{e\,\hbar}{2\,m_e\,c}\,
  \sqrt{\frac{3}{4\,\pi}}\,J_{\mu}
\end{eqnarray}
where $J_{\mu}$ is the atomic angular momentum in the intrinsic frame
\begin{equation}
  J_{\mu}= \,\frac{1}{\hbar}\int\,d{\bf r}\,\rho\,(\,{\bf r}\,)\,
  ({\bf r}\wedge{\bf p})_{\mu}.
\end{equation}
By the rotation $R= e^{-i\,J_{y}\,\theta}e^{-i\,J_z\,\varphi}$
we can transform it to the cell frame
\begin{eqnarray}
  R\,J_{\mu}\,R^{\dagger}\,=\,\left\{ 
    \begin{array}{rl}
      &e^{i\,\mu\,\varphi}\biggl(\,i\cot\theta\,\frac{\partial}{\partial\,\varphi}
       +\,\mu \,\frac{\partial}{\partial\,\theta}\,\biggr),\,\,\,\,\,\mu=\pm1 \\
      &-i\,\frac{\partial}{\partial\,\varphi},\,\,\,\,\,\,\,\,\,\,\,\,\,\,\,\,
       \,\mu=0\\
    \end{array}
  \right.
\end{eqnarray}
Taking in to account the $r$-symmetry the $M1$ transition amplitude is
\begin{eqnarray}
  &&\<\,\psi_{n,m}\,|\,{\cal M}(M1,\mu)\,|\,\psi_{0,0}\,\>\,=
   \mu \, \frac{e\,\hbar}{2\,m_e\,c}\,\sqrt{\frac{3}{4\,\pi}}\nonumber \\
  &&\times \,\delta_{m\,\mu}\,2\,\<\,\chi_{n,|\mu|}\,|\,
   \frac{\partial}{\partial\,\theta}\,|\,\chi_{0,0}\,\>.
\end{eqnarray}
Performing the change of variables $\xi=\theta^2/\theta^2_0$
\begin{eqnarray}
  &&\<\,\chi_{n,|\mu|}\,|\,\frac{\partial}{\partial\theta}\,|\,\chi_{0,0}\,\>
   =\nu_{n,\,|\mu|}\,\nu_{0,\,0}\,\theta_0\int_0^\infty\,d\,\xi\,\xi \nonumber \\
  &&\times e^{-\,\frac{\xi}{2}}\,L_{n}^{(|\,\mu|)}(\xi)
   \,\frac{\partial}{\partial\,\xi}
   \,\biggl[\,e^{-\,\frac{\xi}{2}}\,L_{0}^{(0)}(\xi)\,\biggr]
\end{eqnarray}
we get its final expression
\begin{eqnarray}
  \<\psi_{n,m}|{\cal M}(M1,\mu)|\psi_{0,0}\>
  =-\mu \frac{e\,\hbar}{m_e c\,\theta_0}\sqrt{\frac{3}{4\,\pi}}
  \delta_{m\mu}\delta_{n0}.
\end{eqnarray}
As anticipated only the states $n=0$, $m=\pm 1$ are excited. The fact
that the amplitude vanishes for $\mu=0$ is due to the fact
that the atom can rotate only around the $z$-axis.

Next we evaluate the $E2$ transition amplitude. The intrinsic quadrupole
moment is
\begin{eqnarray}
  Q_{2\,\mu}\, &=& e \int d{\bf r} \rho({\bf r})\,r^2\,Y_{2\,\mu}(\hat{\bf r}).
  \nonumber
\end{eqnarray}
We parametrize the charge density according to
$\rho({\bf r})= s(r(\theta,\,\varphi)-r)\, \rho_0$
where $s$ is the step function and
$r(\theta,\,\varphi)=r_0\,(1+\alpha_{20}Y_{20}(\hat{\bf r}))$,
so that
\begin{eqnarray}
  Q_{2\,\mu} \simeq e \, \rho_0 \,r_0^5\,\alpha_{20}\,\delta_{\mu 0}.
\end{eqnarray}
Transforming to the cell frame we have
\begin{eqnarray}
  {\cal M}(E2,\mu)=\sum_{\nu}Q_{2\nu}\<2\nu|R|2\,\mu\>
  =Q_{20}\,d^{(2)}_{0\mu}(\theta)e^{-i\mu\varphi},
\end{eqnarray}
where
\begin{eqnarray}
  d^{(2)}_{0\mu}(\theta)=(-1)^{\mu}\sqrt{\frac{(2-\mu)!}
  {(2+\mu)!}}\,P_2^{\mu}(\cos\theta).
\end{eqnarray}
So the $E2$ transition amplitude becomes
\begin{eqnarray}
  &&\<\,\psi_{n,m}\,|\,{\cal M}(E2,\mu)\,|\,\psi_{0,0}\,\>\,
   =\delta_{m\mu}\,Q_{20} \nonumber \\
  &&\hspace{-10mm}\times\int_0^{\infty}\,d\,{\theta}\,\theta
   \chi_{n,|\mu|} [d^{(2)}_{0\,\mu}\,(\theta)+(-)^m\,d^{(2)}_{0\,\mu}
   \,(\pi-\theta)\,]\chi_{0,0} \, \nonumber \\
  &&=2\,\delta_{m\mu}\,Q_{2\,0}\int_0^{\infty}\,d\,{\theta}\,\theta
   \,\chi_{n,|\mu|}\, d^{(2)}_{0\,\mu}\,(\theta)\,\chi_{0,0},
\end{eqnarray}
since $ d^{(2)}_{0\,\mu}\,(\pi-\theta)=(-)^m\,d^{(2)}_{0\,\mu}\,(\theta) $.
The amplitude for excitation of the lowest state is
\begin{eqnarray}
  \<\,\psi_{0,m}\,|\,{\cal M}(E2,\mu)\,|\,\psi_{0,0}\,\>\,
  =\mu \, \sqrt{\frac{3}{2}}\,\theta_0\,Q_{2\,0}\, \delta_{m\mu}.
\end{eqnarray}
Again we find that only the states $m=\pm1$ are excited.

The excitation energy and the electromagnetic transition amplitudes depend only
on the parameters $C$ and $I$. To perform an estimate of these parameters
we must distinguish the ionic from the covalent binding. As a prototype of the
first one we consider LaMnO$_3$. In this crystal the Mn$^{3+}$ ion is
surrounded by six O$^{2-}$ ions sitting at the vertices of an octahedron
which is elongated in the $z$-direction as shown in the Figure

The restoring force originated by the quadrupole-quadrupole
electromagnetic interaction is a function of the rotation angle $\theta$
\begin{eqnarray}
  V(\,\theta\,)= \frac{1}{6}\,\sum_{i}
  \,\<\, \delta q_{ii} \,\>
  \,\<\frac{1}{R^{5}}\,Q_{ii}\>,
\end{eqnarray}
where $ \delta q_{ ii } = q_{ii}(\,\theta\,)-\,q_{ii}(\,0\,) $.
This equation holds if the charge of the ligands is external to the atom.
$Q_{ii}$ are the diagonal components of the electric quadrupole moment
density of the cell (we assume $Q_{ij}=\,0$, $i\,\ne\,j$) and
$q_{ii}(\,\theta\,)$ are the electric quadrupole components of the atomic
electrons after their coordinates are rotated by the angle $ \theta $
through the $x$-axis. The expectation values are taken with respect to the ligand and atomic wave
functions respectively. To second order in $\theta$ we have
$ \delta q_{ 11 } = 0 $, $ \delta q_{ 22 } = - (\,q_{22}-q_{33}\,)\,\theta^2 $,
$ \delta q_{ 33 } = \,(\,q_{22} - q_{33}\,)\,\theta^2 $.
We assume that the atomic charge has an axially symmetric ellipsoidal
shape with axes $ c = r_0\,(\,1+\,2\,\delta\,), \, 
a = r_0\,(\,1-\,\delta\,) $,
while the ligands are point-like with coordinates
\begin{eqnarray}
  &&(0,0,R_0(1+\,2\,\triangle)), (0,0,-R_0(1+\,2\,\triangle)),\nonumber \\
  &&(R_0(1-\triangle),0,0), (-R_0(1-\triangle),0,0),\nonumber \\
  &&(0,R_0(1-\triangle),0), (0,-R_0(1-\triangle),0).
\end{eqnarray}
Therefore the following relations hold
\begin{eqnarray}
  \<\,q_{33}\,\>\, &=&\, -2\,\<\,q_{22}\,\>\, =\, -2\,\<\,q_{11}\,\> \nonumber\\
  \<\,\frac{1}{R^5}\,Q_{33}\,\>\, &= &\,-2\,\<\,\frac{1}{R^5}\,Q_{22}\,\>\,
  = \,-2\,\<\,\frac{1}{R^5}\,Q_{11}\,\>
\end{eqnarray}
and
\begin{eqnarray}
  \<q_{33}\> \simeq \frac{12\,Z_a\,e}{5}\,r_0^2\,\delta,\,\,\,
  \<\,\frac{1}{R^5}\,Q_{33}\,\>\simeq-36\,e\,Z_{l}\,\frac{\triangle}{R_{0}^3}.
\end{eqnarray}
$Z_l$, $Z_a$ are the charges of the ligands and the atom respectively.
Assuming for simplicity $\triangle=\delta$ we get
\be
  C = \frac{324}{5}\,Z_{l}\,Z_{a}\,e^2\,\frac{r_{0}^2}{R_{0}^3}\,\delta^2.
\ee

Since the moment of inertia of the atom, under the simplifying assumption
of constant density in the ellipsis, is
\be
  I_{1} \simeq \frac{2\,m_e\,Z_a}{5}\,r_0^2\,(\,1+\,\delta\,),
\ee
we finally get

\be
  \omega^2\simeq \frac{162\,Z_{l}\,e^2\,\delta^2}{m_e R_{0}^3},\,\,\,
  \theta_0^2 = \frac{5}{18\sqrt{2}\,Z_a\,\delta\,\sqrt{Z_{l}}}
  \frac{R_0^{\frac{3}{2}}}{r_0^2}.
\ee
According to \cite{ravindran}, in LaMnO$_3$ the distortion parameter is
$\triangle = 0.06\,$, while $r_0=1.42\,{\rm \AA}$, $R_{0}$ = $2.02\,{\rm \AA}$,
$Z_{l}$ = 2, Z$_a=22$. With the above values we get
$
  \hbar\omega = 4.0 \, {\rm eV},\,\,\,
  \theta_0 = 0.33 \,\,\,{\rm radiants}.
$
The small value of $ \theta_0$ justifies the extension we did of the integrals
over $ \theta $ up to infinity.
The exact value~\cite{abramowitz} of $\omega$ differs from the above
by less than 6\%.

 A major difference with respect to the nuclear case has already
been emphasized: In standard nuclear experiments the nuclei are unpolarized,
while in crystals the deformed atoms are essentially polarized.
As a consequence the electromagnetic cross sections have an angular dependence
which should make their measurement possible. Indeed the cross section
is maximum when the momentum of the photon is parallel to the anisotropy axis
and vanishes when it is perpendicular to it.
Therefore a differential dichroic absorption with linear polarization of the 
incoming photon parallel and perpendicular to the anisotropy axis should
be able to single out the magnetic dipole transition due to the collective 
scissors mode, provided that around its energy the linear 
dichroism originating from electric dipole ($E1$) transitions be negligible or 
structureless. This latter feature should obviously be derived from a 
realistic description of the continuum particle-hole excitations based on 
the band structure of the crystal. In this connection we note that the damping of 
the collective mode is expected either to be zero, if its energy falls in the 
band gap, or rather small if it falls in the particle-hole continuum, due to the 
smallness of the matrix element connecting the mode to the particle-hole states.

\begin{figure}
  \begin{center}
    \begin{tabular}{cc}
      \includegraphics[width=4cm]{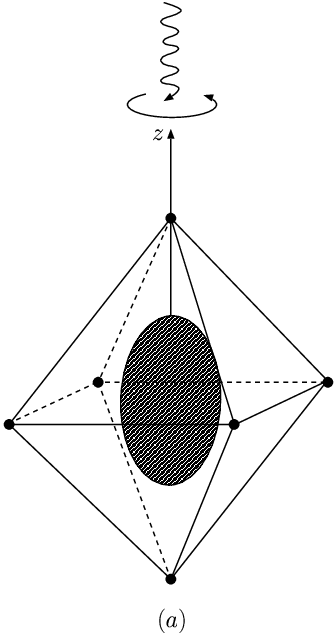}
      \includegraphics[width=4cm]{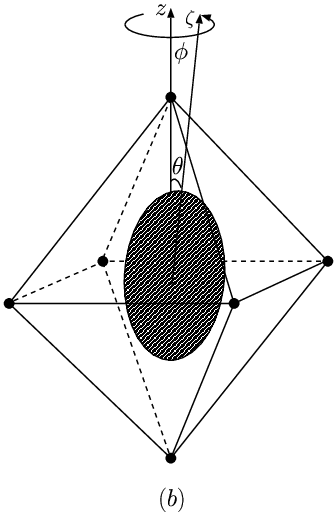}
    \end{tabular}
  \end{center}
  \caption{ In (a) the atom is in the ground state and the incoming photon
    has circular polarisation. In (b) the atom
    carries the angular momentum of the absorbed photon and precesses at an angle
    $\theta_0$ around the cell axis generating a magnetic moment proportional
    to $1 / \theta_0$, with a variation of quadrupole moment proportional to $\theta_0$. }
  \label{f1}
\end{figure}

In a triaxial deformation an energy splitting of the collective mode is 
expected, as predicted in nuclei~\cite{Palu}. Such a splitting might obviously provide 
additional pieces of information, but in a more realistic microscopic 
description based on the Random Phase Approximation, as done 
in Nuclear Physics, one should also expect a fragmentation of the mode that 
might mask it and make its experimental detection more difficult.

As a last remark we may note that inelastic photon scattering is another 
technique able to detect the scissors mode.

The case of covalent binding cannot be analyzed in the same way. First, the
decoupling of the intrinsic degrees of freedom from the collective (rotational)
ones, which is at the basis of our semiclassical treatment, is less obvious.
Second, the estimate of the restoring force and of the moment of inertia
requires a numerical calculation. Indeed we expect that the restoring force will result smaller.
 Yet we deem that the scissors mode will
not be precluded by the nature of the covalent binding, and that the above
numerical estimates, should still remain valid as an order of magnitude.

\subsection *{Acknowledgment}

We are grateful to M. Benfatto and C. R. Natoli for many fruitful discussions.

\end{document}